\documentclass[%
reprint,
superscriptaddress,
amsmath,amssymb,
aps,
pra,
letterpaper,
]{revtex4-1}
\usepackage{graphicx}
\usepackage{mathtools} 
\usepackage{textcomp} 
\usepackage[space]{grffile} 
\usepackage{dcolumn}
\usepackage{bm}
\usepackage{textgreek} 
\usepackage{hyperref}
\usepackage{epstopdf}

\begin{document}

\title{Optical coherence of diamond nitrogen-vacancy centers formed by ion implantation and annealing}


 \author{S. B. van Dam}
 \thanks{These authors contributed equally.}
 \affiliation{QuTech, Delft University of Technology, PO Box 5046, 2600 GA Delft, The Netherlands}
 \affiliation{Kavli Institute of Nanoscience, Delft University of Technology, PO Box 5046, 2600 GA Delft, The Netherlands}

 \author{M. Walsh}
 \thanks{These authors contributed equally.}
\email{mpwalsh@mit.edu}
 \affiliation{Department of Electrical Engineering and Computer Science, Massachusetts Institute of Technology, Cambridge, Massachusetts 02139, USA}

 \author{M. J. Degen}
 \affiliation{QuTech, Delft University of Technology, PO Box 5046, 2600 GA Delft, The Netherlands}
 \affiliation{Kavli Institute of Nanoscience, Delft University of Technology, PO Box 5046, 2600 GA Delft, The Netherlands}

 \author{E. Bersin}
 \affiliation{Department of Electrical Engineering and Computer Science, Massachusetts Institute of Technology, Cambridge, Massachusetts 02139, USA}

 \author{S. L. Mouradian}
 \thanks{Present Address: Department of Physics, University of California Berkeley, California 94720, USA}
 \affiliation{Department of Electrical Engineering and Computer Science, Massachusetts Institute of Technology, Cambridge, Massachusetts 02139, USA}

 \author{A. Galiullin}
 \affiliation{QuTech, Delft University of Technology, PO Box 5046, 2600 GA Delft, The Netherlands}
 \affiliation{Kavli Institute of Nanoscience, Delft University of Technology, PO Box 5046, 2600 GA Delft, The Netherlands}

 \author{M. Ruf}
 \affiliation{QuTech, Delft University of Technology, PO Box 5046, 2600 GA Delft, The Netherlands}
 \affiliation{Kavli Institute of Nanoscience, Delft University of Technology, PO Box 5046, 2600 GA Delft, The Netherlands}

 \author{M. IJspeert}
 \affiliation{QuTech, Delft University of Technology, PO Box 5046, 2600 GA Delft, The Netherlands}
 \affiliation{Kavli Institute of Nanoscience, Delft University of Technology, PO Box 5046, 2600 GA Delft, The Netherlands}

 \author{T. H. Taminiau}
 \affiliation{QuTech, Delft University of Technology, PO Box 5046, 2600 GA Delft, The Netherlands}
 \affiliation{Kavli Institute of Nanoscience, Delft University of Technology, PO Box 5046, 2600 GA Delft, The Netherlands}

 \author{R. Hanson}
 \email{r.hanson@tudelft.nl}
 \affiliation{QuTech, Delft University of Technology, PO Box 5046, 2600 GA Delft, The Netherlands}
 \affiliation{Kavli Institute of Nanoscience, Delft University of Technology, PO Box 5046, 2600 GA Delft, The Netherlands}

 \author{D. R. Englund}
 \affiliation{Department of Electrical Engineering and Computer Science, Massachusetts Institute of Technology, Cambridge, Massachusetts 02139, USA}


\begin{abstract}
The advancement of quantum optical science and technology with solid-state emitters such as nitrogen-vacancy (NV) centers in diamond critically relies on the coherence of the emitters' optical transitions. 
A widely employed strategy to create NV centers at precisely controlled locations is nitrogen ion implantation followed by a high-temperature annealing process.  
We report on experimental data directly correlating the NV center optical coherence to the origin of the nitrogen atom. 
These studies reveal low-strain, narrow-optical-linewidth ($<500$~MHz) NV centers formed from naturally-occurring $^{14}$N atoms.
In contrast, NV centers formed from implanted $^{15}$N atoms exhibit significantly broadened optical transitions ($>1$~GHz) and higher strain.
The data show that the poor optical coherence of the NV centers formed from implanted nitrogen is not due to an intrinsic effect related to the diamond or isotope. 
These results have immediate implications for the positioning accuracy of current NV center creation protocols and point to the need to further investigate the influence of lattice damage on the coherence of NV centers from implanted ions.
\end{abstract}
\maketitle


Coherent optical control over solid-state quantum emitters has enabled new advances in quantum science \citep{Aharonovich2016,Awschalom2018,Lodahl2018} and may lead to technologies such as quantum networks \citep{Wehner2018}. A quantum network crucially relies on entanglement connections that can be established through a coherent spin-photon interface. The nitrogen-vacancy (NV) defect center in diamond is a well-suited candidate owing to a spin ground state with a long coherence time \citep{Bar-Gill2013, Abobeih2018}, nearby nuclear spins for quantum memories \citep{Kalb2017} or algorithms \citep{Waldherr2014,Cramer2016,Lovchinsky2016}, and spin-selective optical transitions allowing for efficient optical spin initialization and readout \citep{Robledo2011}. Moreover, at low strain and low temperature ($< 10$~K), effects from phonon mixing in the excited state are small \citep{Fu2009,Goldman2015}, and the optical transition can be coherent. Indeed, narrow-linewidth, coherent optical transitions \citep{Tamarat2006, Batalov2008, Robledo2010} have been used for the generation of indistinguishable photons suited for two photon quantum interference \citep{Bernien2012, Sipahigil2012} and entanglement generation between remote NV centers \citep{Bernien2013}.

To date, all experiments employing coherent photons from NV centers have been performed with NV centers that were formed during diamond growth. Key to their optical coherence is that these NV centers experience an environment with few defects, since the stability of optical transitions (as with many solid-state systems) suffers from unwanted interactions with nearby bulk and surface defects leading to changes in the strain and electric field environment \citep{Manson2006,Fu2010,Orwa2011,Santori2010,Faraon2012}. For NV centers with a broadened linewidth below $\approx200$~MHz dominated by slow spectral diffusion, protocols using resonant charge repumping \citep{Siyushev2013} and real-time monitoring of the transition frequency \citep{Hensen2015} have been used to reduce the broadened linewidth to an effective linewidth of below $50$~MHz, suitable for quantum optical experiments. However, such protocols are challenging for NV centers with greater spectral diffusion.

Instead of being limited to NV centers formed during diamond growth, they can be created, for example by nitrogen ion implantation \citep{Pezzagna2011a}. Nitrogen ion implantation provides an NV positioning accuracy that enables integration with on-chip photonics \citep{Mouradian2015, Schroder2016} and coupling between nearby NV centers \citep{Gaebel2006, Dolde2013, Yamamoto2013}. Precise positioning of NV centers or accurately registering their location is also a prerequisite for optimal overlap of the dipole with the electric field mode of diamond optical cavities, for engineering and enhancing light-matter interaction \citep{Faraon2011, Faraon2012, Hausmann2013, Lee2014, Li2015, Riedrich-Moller2015, Riedel2017}. Moreover, ion implantation allows for the creation of single NV centers in high purity diamond, providing a potentially low-defect environment \citep{Orwa2011}.
 
\begin{figure}
\includegraphics[width=\linewidth]{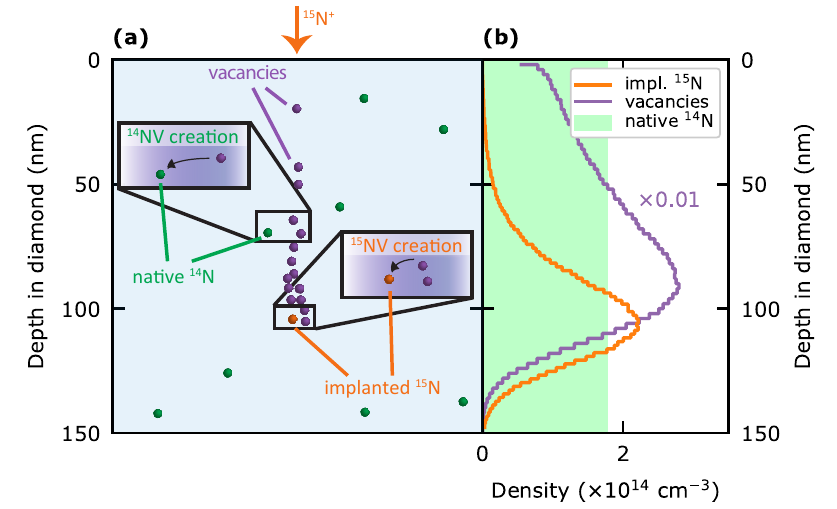}
\caption{\textbf{NV creation via nitrogen-ion implantation.} \textbf{a}, Schematic showing implanted $^{15}$N$^+$ ions (orange) leaving a trail of vacancies (purple) until settling into a final position. Naturally abundant $^{14}$N ions (green) are shown randomly distributed throughout the diamond lattice. Vacancies (implantation-induced or native) mobilized by annealing can bind to a nitrogen atom (implanted or native). \textbf{b}, A SRIM simulation using the parameters in sample A show the distribution of implanted nitrogen (orange) and created vacancies (purple). The shaded green area indicates the range of the estimated natural $^{14}$N concentration reported by Element 6.}
\label{fig:1}
\end{figure}

However, the bombardment of the diamond with nitrogen ions creates crystal damage that can deteriorate spin and optical coherence properties of NV centers \citep{Fu2010,Orwa2011}. High-temperature annealing can mitigate some of these issues by repairing the diamond lattice \citep{Lea-Wilson1995,Naydenov2010,Yamamoto2013a,Deak2014}. A procedure including a low implantation dose, careful cleaning, and high-temperature annealing was reported by Chu et al. \citep{Chu2014}, leading to the creation of narrow-linewidth NV centers. These narrow-linewidth NV centers can result from implanted nitrogen atoms, or from native nitrogen atoms, combined with for example implantation-induced vacancies (Figure \ref{fig:1}a). 
In principle, the source of nitrogen can be verified by implanting $^{15}$N isotopes (natural abundance 0.37\%) and resolving the hyperfine structure of the NV magnetic spectrum \citep{Rabeau2006}, as done in studies of the spin coherence \citep{Ofori-Okai2012,Yamamoto2013a} and creation efficiency \citep{Gaebel2006,Naydenov2010a,Toyli2010,Yamamoto2014, Becker2018} of NVs formed from implanted nitrogen. However, in Chu et al. \citep{Chu2014} the isotope of the narrow-linewidth NV centers was not investigated \citep{Chu_comm}. In a later study with similar results \citep{Riedel2017}, $^{14}$N isotopes were implanted, so that the origin of the NV center's nitrogen atom could not be determined.
Here we report on a study that enables us to directly correlate the optical linewidth of NV transitions to the NV formation mechanism. 

To distinguish NVs formed by implanted nitrogen atoms from those formed by native nitrogen atoms, we implanted $^{15}$N isotopes \citep{Rabeau2006}. We then experimentally correlated the optical linewidth to the nitrogen isotope. The study was carried out on two separate samples~\cite{supplement}. Sample A (processed at MIT) is a bulk $\left<100\right>$ CVD grown diamond (Element 6), prepared with the same implantation and annealing procedure as presented in Chu et al. \citep{Chu2014}: it was implanted with $^{15}$N$^+$ at 85~keV with a fluence of $10^9$~N/cm$^2$ and subsequently annealed at a maximum temperature of $1200$~$^{\circ}$C. Sample B (Delft) is a membrane (thickness $\approx14$~\textmu m) obtained from a bulk $\left<100\right>$ CVD grown diamond (Element 6), implanted with $^{15}$N$^+$ at 400~keV (fluence $10^8$~N/cm$^2$), and subsequently annealed at a maximum temperature of $1100$~$^{\circ}$C.
 
During implantation, nitrogen ions penetrate the diamond to a depth determined by the implantation energy (Figure~\ref{fig:1}). As implanted nitrogen atoms track through the crystal, they displace carbon atoms from their lattice sites creating vacancies. The nitrogen atoms create damage along the entire trajectory, but the damage is greatest near the stopping point \citep{DeOliveira2017}. We performed SRIM \cite{SRIM} simulations to predict the stopping point of implanted $^{15}$N atoms, in addition to the locations of vacancies created along the trajectory (Figure~\ref{fig:1}b). At temperatures $>600$~$^{\circ}$C, vacancies become mobile \citep{Davies1992}. These vacancies can form an NV center, recombining with the implanted $^{15}$N that created the damage or with a native $^{14}$N in the lattice. 
The resulting $^{15}$NV and $^{14}$NV formation yields can vary significantly \citep{Gaebel2006,Naydenov2010a,Toyli2010,Ofori-Okai2012,Yamamoto2014} depending on several factors, including the initial nitrogen concentration, the implantation fluence and energy, the number of vacancies created during the implantation process, and the duration and temperature of annealing.

\begin{figure*}
\includegraphics[trim={1in, 0, 0.375in, 0}]{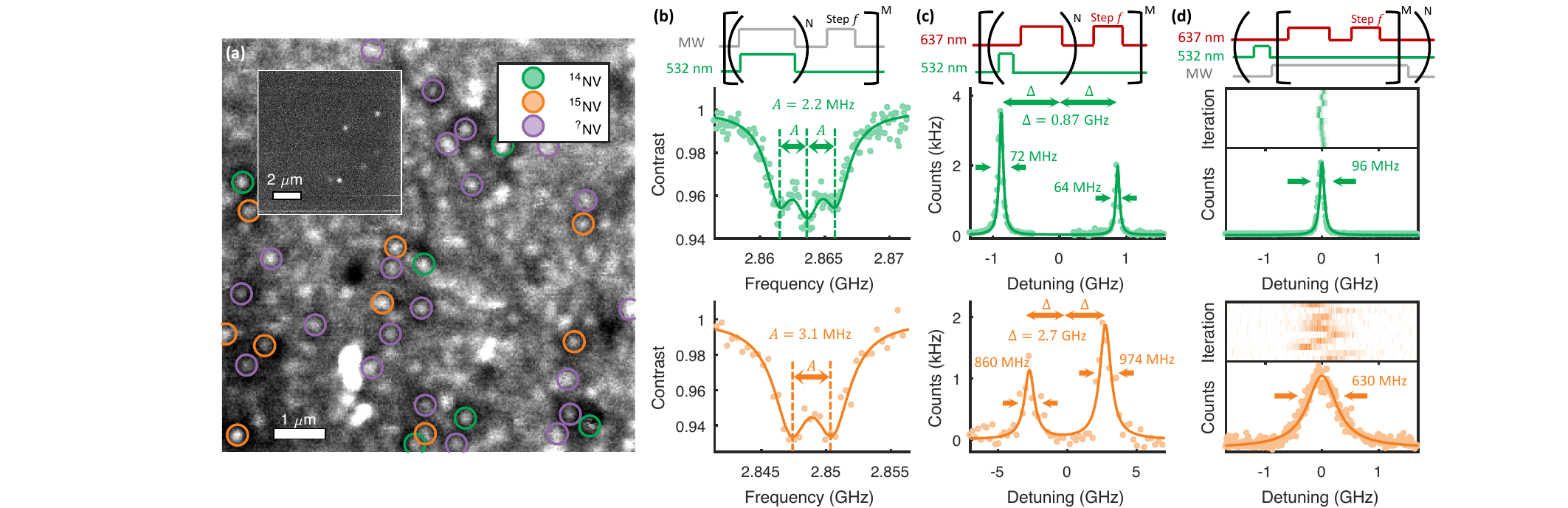}
\caption{\textbf{Isotope characterization and optical measurements of NV centers from $^{14}$N and $^{15}$N.} \textbf{a}, A fluorescent confocal scan of sample A taken at 4~K, with labels indicating NV centers characterized as $^{14}$NV, $^{15}$NV, and a set with unresolvable hyperfine lines, labeled as $^?$NV. A scan a few microns below the implanted layer (inset) shows a lower NV density. \textbf{b-d}, Pulse sequences (top row) used for isotope characterisation and optical measurements, and representative measurement results for each isotope (green, middle row: $^{14}$NV, orange, bottom row: $^{15}$NV). \textbf{b}, Continuous wave (CW) ODMR measurements reveal the NV isotope. The $^{14}$NV is characterized by $S=1$ hyperfine transitions; the $^{15}$NV by $S=1/2$ hyperfine transitions. \textbf{c}, Interleaved red and green excitation probe the combined effect of short-timescale fluctuations and laser-induced spectral diffusion. The $E_x$ and $E_y$ ZPL transitions are visible for both isotopes; the $^{14}$NV linewidths are narrower and show a smaller strain splitting than the $^{15}$NV. \textbf{d}, Individual line scans of the ZPL in sample B reveal the linewidth free from laser-induced spectral diffusion. The summation of many repeated scans is broadened as a result of repump-laser-induced spectral diffusion.}
\label{fig:2}
\end{figure*}

A representative confocal fluorescence map at the implantation depth in sample A is shown in Figure~\ref{fig:2}a. Confocal fluorescence scans at foci deeper into the diamond show a significantly lower density of fluorescent spots (Figure~\ref{fig:2}a, inset), indicating that the emitters near the surface were predominantly created by the implantation and annealing process \citep{supplement}. We identified emitters using different protocols in the two samples.
In sample A, automated spot-recognition was performed on a fluorescence scan. For each detected spot we identified an NV center based on its characteristic zero-phonon line (ZPL) emission around 637~nm using a spectrograph from a photoluminescence measurement at 4 K under 532~nm excitation. This protocol identified 120 fluorescent spots as NV centers in a $\approx400$~\textmu m$^2$ area.
In sample B, spots in a fluorescence scan were detected visually, after which an automated protocol identified NV centers based on the presence of a resonance in an optically detected magnetic resonance (ODMR) spectrum around the characteristic NV center zero-field splitting of 2.88~GHz. In this way, 52 out of a total 57 inspected spots in a $\approx75$~\textmu m$^2$ area in the implantation layer were identified as NV centers.

We next determined the nitrogen isotope of each NV center by observing the hyperfine structure of the ODMR spectra. A weak external magnetic field (B$_\parallel$ $\approx$ 5-10~G) was applied to separate the $m_s=-1$ and $m_s=+1$ electron spin transitions. We found NV centers with the characteristic triplet splitting of the $^{14}$NV (with hyperfine splitting, $A=2.2$~MHz) as well as with the $^{15}$NV doublet ($A=3.1$~MHz) \citep{Doherty2013}, as indicated in Figure~\ref{fig:2}b. Of the 120 NVs identified on sample A, an ODMR signal was detected in 50, out of which 18 were $^{15}$NV, 18 $^{14}$NV, and there were 14 in which the isotope could not be reliably determined from the ODMR spectra. Similarly, of the 52 NVs identified on sample B, 34 were $^{15}$NV, 3 were $^{14}$NV, and the isotope could not be determined in 15 NVs. We attribute the different isotope occurrence ratios in sample A and B to different native $^{14}$N content and different implantation fluence. 

Subsequently, we measured the linewidth of optical transitions of identified NV centers, recording photoluminescence excitation (PLE) spectra at low temperature ($\approx 4$~K). 
A tunable laser with a wavelength near 637~nm was scanned over the optical transition while detecting emitted photons in the phonon-sideband. We performed two types of measurements. 
First, a scan was made in which resonant excitation (637~nm) and green illumination (532~nm) were rapidly interleaved at each data point. The red excitation causes rapid optical spin pumping and ionization of the NV center. The green excitation provides repumping into the negative charge state and the $m_s=0$ spin state.  This measurement reveals the combined effect of short-time scale fluctuations and repump-laser-induced spectral diffusion in broadening the transition linewidth. Examples of the resulting traces are seen in Figure~\ref{fig:2}c.

\begin{figure*}
\includegraphics{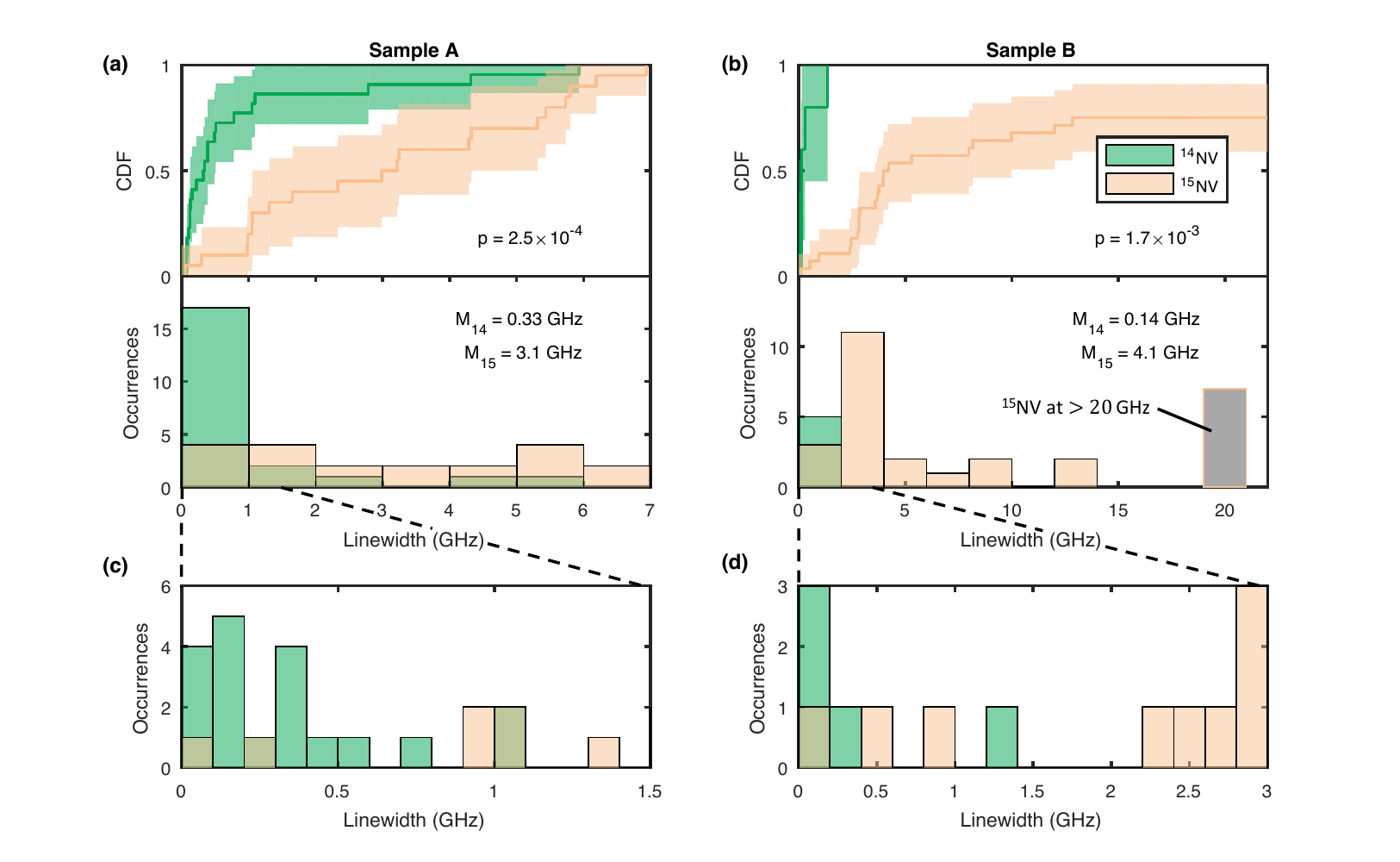}
\caption{\textbf{Optical linewidths per isotope.} \textbf{a-b}, A summary of the optical linewidths identified in sample A (\textbf{a}) and sample B (\textbf{b}) from scans at the implantation depth. For sample B, that has comparatively few $^{14}$NV centers at the implantation depth, we included three $^{14}$NV centers found deeper in the diamond to enable a comparison between NVs formed from implanted versus native nitrogen. The distribution is represented as a cumulative distribution function (CDF, top), with the corresponding histogram shown below. The shaded region in the CDF indicates a 95\% confidence interval calculated using Greenwood's formula. These data show that both diamonds supported narrow-linewidth NV centers, the majority of which originated from $^{14}$NVs. While $^{15}$NV do exhibit narrow lines, their median linewidth (M) is higher than for the $^{14}$NV centers in both samples. We evaluate the probability to obtain the observed linewidths for $^{14}$NV centers and $^{15}$NV centers if the samples are drawn from the same distribution with a Wilcoxon Rank Sum test, finding a p-value of $2.5\times10^{-4}$ in sample A and $1.7\times10^{-3}$ in sample B. \textbf{c-d}, A magnification of the histograms shown in \textbf{a} and \textbf{b}.  }
\label{fig:3} 
\end{figure*}

Second, in sample B, an additional scan was performed to isolate the effect of short timescale fluctuations from repump-laser-induced diffusion. A single off-resonant repump was applied before sweeping the resonant laser at low power, as seen in Figure~\ref{fig:2}d. We applied microwaves on the spin resonances to prevent optical pumping into a dark spin state during the sweeps \citep{Tamarat2008}. 
Remaining traces in which the NV center ionized were excluded by applying a second scan over the resonance to check the charge state. If no resonance was observed, the preceding trace was disregarded. This scanning protocol was repeated many times to probe spectral diffusion through the resulting spread of the observed lines \citep{Fu2009}. To extract the linewidth free from repump-laser-induced spectral diffusion, we performed a weighted average of linewidth values found from Lorentzian fits to each individual scan.
 
Figures~\ref{fig:2}c and d display representative resonant optical scans for the $^{14}$NV and $^{15}$NV centers, each showing two resonances corresponding to the two $m_s=0$ orbital transitions $E_x$ and $E_y$. Notably, while the $^{14}$NV center (green, top row) exhibits a narrow optical linewidth with a full-width-at-half-maximum (FWHM) of $64\pm4$~MHz, the $^{15}$NV linewidth (orange, bottom row) is broad, with a FWHM of $860\pm236$~MHz. The dynamics in the second scan type (Figure \ref{fig:2}d) indicate that both repump-induced fluctuations and a short-timescale mechanism broaden the $^{15}$NV linewidth, but that repump-induced fluctuations are dominant in broadening beyond 200~MHz \citep{supplement}.

To correlate the occurrence of narrow optical linewidths with the N isotope of the NV centers, we acquired an extensive data set using the data accumulation procedures described above. The resulting distributions of optical linewidths for both N isotopes are shown in Figure~\ref{fig:3}. Narrow optical linewidths in both samples can be attributed almost exclusively to NVs with a native $^{14}$N host. In contrast, $^{15}$NV centers exhibiting narrow optical linewidths are extremely rare, with a median linewidth for $^{15}$NV centers of 3.1~GHz in sample A and 4.1~GHz in sample B. 

\begin{figure*}
\includegraphics[width=\linewidth]{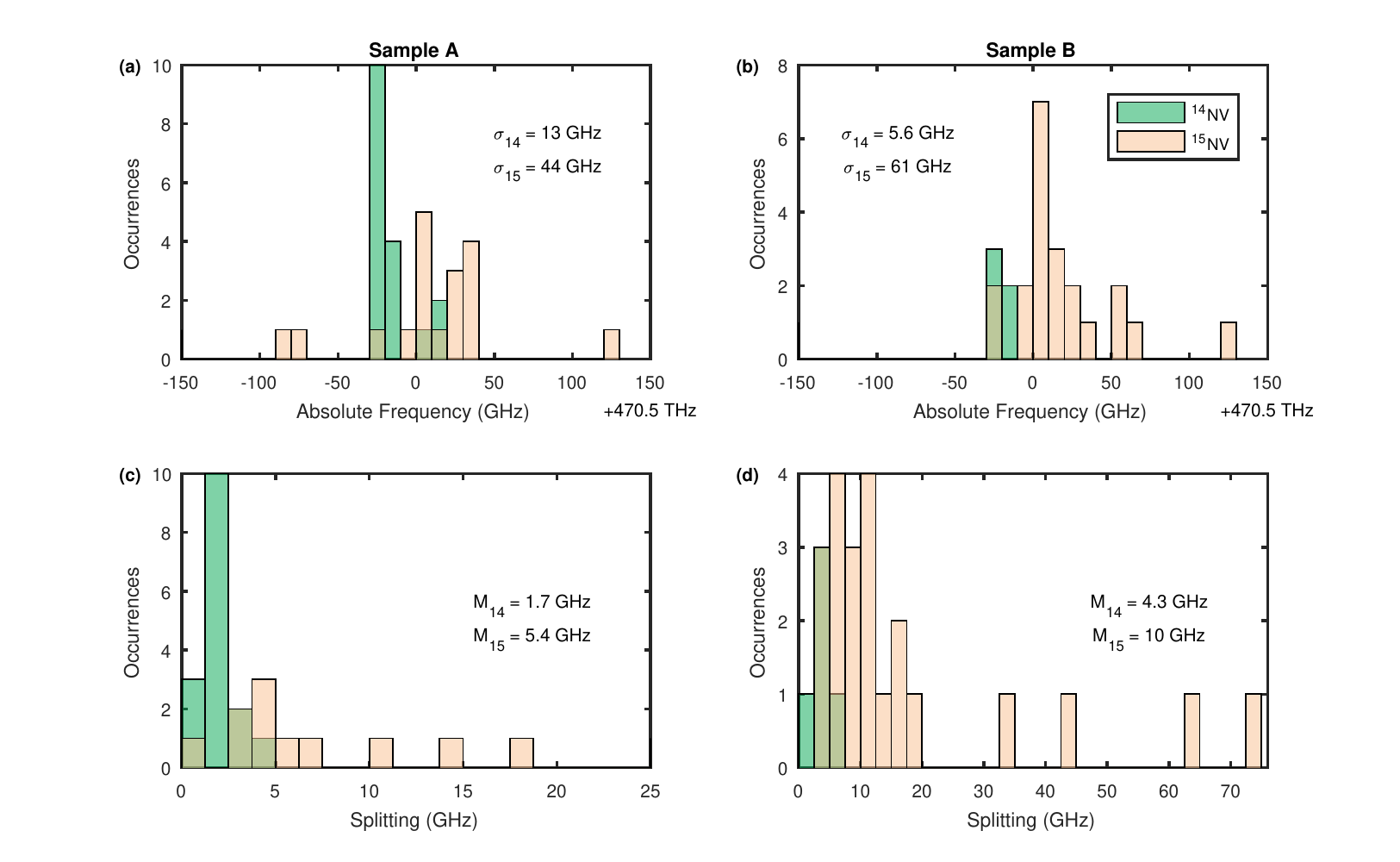}
\caption{\textbf{Strain analysis.}  \textbf{a-b}, The distribution of axial strain (measured by absolute average ZPL frequency) in NVs acquired from analysis of sample A (\textbf{a}) and sample B (\textbf{b}). The $^{15}$NV ZPLs exhibit a larger spread in axial strain (standard deviation, $\sigma$) than the $^{14}$NV ZPLs. \textbf{c-d}, The distribution of transverse strain (measured by half the splitting between $E_x$ and $E_y$ frequencies) in NV centers of sample A (\textbf{c}) and sample B (\textbf{d}). The $^{15}$NV ZPLs show a greater median splitting (M) in both samples. }
\label{fig:4} 
\end{figure*}

Notably, in both datasets one NV center with a $^{15}$N host was found that showed narrow optical linewidths ($< 100$~MHz). 
Given their low occurrence and the non-zero natural abundance of $^{15}$N, the creation mechanism of these narrow-linewidth $^{15}$NVs cannot be conclusively determined.
Nevertheless, their presence demonstrates that $^{15}$NV centers can exhibit coherent optical transitions. Therefore, we conclude that the difference in distribution of optical linewidths between $^{14}$NVs and $^{15}$NVs is not due to an intrinsic effect related to the isotope itself, but due to differences in the local environment resulting from the implantation process.

Damage due to implantation may cause local strain fields. Axial strain results in an overall shift of the optical transition, while transverse strain will split the $E_x$ and $E_y$ transitions \citep{Doherty2013}. The distributions characterizing the strain for both NV isotopes are shown in Figure~\ref{fig:4}. The spread of the distribution in ZPL detuning representing axial strain for $^{15}$NVs (44~GHz for sample A, 60~GHz for sample B) is wider than for $^{14}$NVs (13~GHz for sample A, 5.6~GHz for sample B). Further, we found that $^{15}$NVs exhibit higher transverse strain, manifested by greater splitting with a median of 5.4~GHz (10~GHz) compared to 1.7~GHz (4.3~GHz) for $^{14}$NVs in sample A (B). Assuming a similar strain susceptibility for both isotopes, these results indicate that local damage around the implanted $^{15}$NVs creates a more strained environment, providing further evidence that implantation-induced local damage is responsible for the broadened $^{15}$NV linewidth. In addition, in both samples we observed a shift of the average ZPL frequency for $^{15}$NV compared to $^{14}$NV ZPLs, possibly due to an intrinsic dependency of the energy levels on the isotope as observed in other color centers \citep{Dietrich2014,Ekimov2015}.

The data show indications of an increase in $^{14}$NV density in the implantation layer in both samples \citep{supplement}. We hypothesize that these $^{14}$NV centers can be formed from naturally occurring nitrogen combining with vacancies created during implantation. Since they can be at greater distance from the main damage center near the stopping point of the nitrogen, these NV centers may be coherent and useable for quantum information purposes; more work is needed for a statistically significant correlation. However, these NVs would have worse positioning accuracy as their spatial distribution is set by arbitrarily positioned naturally occurring nitrogen in combination with the diffusion length of the vacancies generated during implantation.

In summary, the implanted nitrogen atoms yield NV centers with predominantly broad optical lines ($> 1$ GHz) and substantially higher strain than NV centers formed from native nitrogen. 
These results indicate that
implanted nitrogen atoms combined with an annealing process at high temperatures do not routinely produce NV centers with narrow optical linewidths.
Vacancies produced in the implantation process may combine with existing nitrogen atoms to produce narrow NVs, but more work is needed for a statistically significant correlation.  
It is clear from this work that recipes for generating implanted NV centers should be re-investigated, addressing local lattice damage associated with implanted nitrogen. 
In addition, other approaches for precisely controlling the NV centers' positions while causing minimal local damage can be further explored, such as employing 2D nitrogen-doped diamond layers combined with electron irradiation or ion implantation for vacancy production \citep{Ohno2012, Ohno2014} or laser writing strategies for creating vacancies with 3D accuracy \citep{Chen2017,Chen2018}.



\begin{acknowledgements}
The authors would like to thank L. Childress, Y. Chu, M. Lon\v{c}ar, P. Maletinsky, M. Markham, M. Trusheim, and J. Wrachtrup for helpful discussions, S. Meesala, and P. Latawiec for sample annealing at Harvard Unversity, and S. Bogdanovic and M. Liddy for help with sample preparation. 
This work was supported in part by the AFOSR MURI for Optimal Measurements for Scalable Quantum Technologies (FA9550-14-1-0052) and by the AFOSR program FA9550-16-1-0391, supervised by Gernot Pomrenke. Further, we acknowledge support by the Netherlands Organization for Scientific Research (NWO) through a VIDI grant, a VICI grant, and through the Frontiers of Nanoscience program, the European Research Council through a Consolidator Grant (QNETWORK) and a Synergy Grant (QC-LAB), and the Royal Netherlands Academy of Arts and Sciences (KNAW) and Ammodo through an Ammodo science award.
M.W. was supported in part by the STC Center for Integrated Quantum Materials (CIQM), NSF Grant No. DMR-1231319, in part by the Army Research Laboratory Center for Distributed Quantum Information (CDQI), and in part by Master Dynamic Limited. E.B. was supported by a NASA Space Technology Research Fellowship and the NSF Center for Ultracold Atoms (CUA). S.L.M. was supported in part by the NSF EFRI-ACQUIRE program Scalable Quantum Communications with Error-Corrected Semiconductor Qubits and in part by the AFOSR Quantum Memories MURI. 

\end{acknowledgements}

\end{document}